\documentclass[12pt]{article}
\usepackage{amsmath}
\usepackage{amssymb}
\newtheorem{theorem}{Theorem}[section]
\newtheorem{lemma}[theorem]{Lemma}

\newtheorem{corollary}[theorem]{Corollary}

\numberwithin{equation}{section}

\begin{document}

\title{SELF-ENERGY OF ELECTRONS IN NON-PERTURBATIVE
QED\thanks{\copyright 1999 by the authors. Reproduction
of this work, in its entirety, by any means, is permitted for
non-commercial purposes by any
means is permitted for non-commercial purposes.}\thanks
{This is a preliminary report on our work presented at the 
University of Alabama,Birmingham - Georgia
Tech International Conference on Differential Equations and Mathematical
Physics, Birmingham, March 15-19, 1999}}

\author{Elliott H. Lieb\thanks{Supported in part
by NSF Grant PHY-98 20650.}\\
Departments of Mathematics and Physics\\
Princeton University, Princeton, New Jersey 08544-0708\\ {\it
lieb@math.princeton.edu}
\and
Michael Loss\thanks{Supported in part by NSF Grant 
DMS-95 00840.}\\ School of Mathematics\\
Georgia Institute of Technology, Atlanta, Georgia 30332-0160 \\ {\it
loss@math.gatech.edu}}

\date{September 15, 1999}
\maketitle
%    General info
%\subjclass{Primary 81V10, 81V70; Secondary 81V45, 81T08, 81T16}

\begin{abstract}
Various models of charged particles interacting with a quantized,
ultraviolet cutoff radiation field (but not with each other) are
investigated. Upper and lower bounds are found for the self- or ground
state-energies without mass renormalization. For $N$ fermions the
bounds are proportional to $N$, but for bosons they are sublinear,
which implies `binding', and hence that `free' bosons are never free.
Both `relativistic' and non-relativistic kinematics are considered. Our
bounds are non-perturbative and differ significantly from the
predictions of perturbation theory.  
\end{abstract}

%%%%%%%%%%%%%%%%%%%%%%%%%%%%%%%%%%%%%%%%%%%%%%%%
\section{Introduction}
\label{intro}

Quantum electrodynamics (QED), the theory of electrons interacting with
photons (at least for small energies) is one of the great successes of
physics.  Among its major achievements is the explanation of the Lamb
shift and the anomalous magnetic moment of the electron.  Nevertheless,
its computations, which are entirely based on perturbation theory,
created some uneasiness among the practitioners.  The occurrence of
infinities was and is especially vexing. Moreover, a truly nontrivial,
3+1-dimensional example of a relativistically invariant field theory
has not yet been achieved.

There are, however, unresolved issues at a much earlier stage of QED
that hark back to black-body radiation, the simplest and historically
first problem involving the interaction of matter with radiation.  The
conceptual problems stemming from black-body radiation were partly
resolved by quantum mechanics, i.e., by the the non-relativistic
Schr\"odinger equation, which  is, undoubtedly, one of the most
successful of theories, for it  describes matter at low energies almost
completely.  It is mathematically consistent and there are techniques
available to compute relevant quantities. Moreover, it allows us to
explain certain facts about bulk matter such as its stability, it
extensivity, and the existence of thermodynamic functions. What has not
been as successful, so far, is the incorporation of radiation phenomena,
the very problem quantum mechanics set out to explain.

It ought to be possible to find a mathematically consistent theory,
free of infinities, that describes the interaction of non-relativistic
matter with radiation at moderate energies, such as atomic binding
energies.  It should not be necessary, as some physicists believe, to
embed QED as a low energy part of a consistent high
energy theory.

{}From such a theory one could learn a number of things that have not
been explained rigorously. i) The decay of excited states in atoms.
This problem has been investigated in some ultraviolet cutoff models in
\cite{BFS} and in a massive photon model in \cite{OY}. See also the review of
Hogreve \cite{H}.
ii) Non-relativistic QED could be a playground for truly non-perturbative
calculations and it could shed light on renormalization procedures. In
fact, this was the route historically taken by Kramers that led to the
renormalization program of Dyson, Feynman, Schwinger and Tomonaga. iii)
Last but not least, one could formulate and answer the problems of
stability of bulk matter interacting with the radiation field.

It has been proved in \cite{F},\cite{LLS} that stability of
non-relativistic matter (with the Pauli Hamiltonian) interacting with
classical magnetic fields holds provided that the fine-structure
constant, $\alpha = e^2/\hbar c$, is small enough. It is certain, that
the intricacies and difficulties of this classical field model will
persist and presumably magnify in QED.

The same may be expected from a relativistic QED since replacing the
Pauli Hamiltonian by a Dirac operator leads to a similar requirement on
$\alpha$ \cite{LSS}. Indeed, stability of matter in this model (the
Brown-Ravenhall model) requires that the electron (positron) be defined
in terms of the positive (negative) spectral subspace of the Dirac
operator {\it with} the magnetic vector potential $A(x)$, instead of
the free Dirac operator without $A(x)$. This observation, that
perturbation theory, if there is one, must start from the dressed
electrons rather than the electrons unclothed by its magnetic field,
might ultimately be important in a non-perturbative QED.

The first, humble step is to understand electrons that interact with
the radiation field but which are free otherwise. In order for this
model to make sense an ultraviolet cutoff has to be imposed that limits
the energy of photon modes.  The simplest question, which is the one
we address in this paper, is the behavior of the self-energy of the
electron as the cutoff tends to infinity (with the bare mass of the
electron fixed). The self-energy of the electron diverges as the cutoff
tends to infinity and it has to be subtracted for each electron in any
interacting theory. The total energy will still depend strongly on the
cutoff because of the interactions. This dependence will, hopefully,
enter through an effective mass which will be set equal to the physical
mass (mass renormalization). The resulting theory should be essentially
Schr\"odinger's mechanics, but  slightly modified by so-called
radiative corrections.

Lest the reader think that the self-energy problem is just a
mathematical exercise, consideration of the many-body problem will
provide a counterexample. Imagine $N$ charged bosons interacting with
the radiation field, but neglect any interaction among them such as the
Coulomb repulsion. We say that these particles bind if the energy of
the combined particles is less than the energy of infinitely separated
particles. As we shall show, charged bosons indeed bind and they do it
in such a massive way that it will be very likely that this cannot be
overcome by the Coulomb repulsion. In particular, the energy of a
charged many-boson system is {\it not} extensive, and from this
perspective it is fortunate that stable, charged bosons do not exist in
nature.

The situation is very different for fermions. We are not able to show that
they do not bind  but we can show --- and this is one of the main results
of our paper --- that the self-energy is extensive, i.e., bounded above
and below by a constant times $N$.

We thus have strong evidence that there is {\it no} consistent
description of a system of stable charged relativistic or
non-relativistic bosons interacting with the radiation field, while the
Pauli exclusion principle, on the other hand, is able to prevent the
above mentioned pathology.

In the remainder of the section we explain our notation and state the
results.  In the subsequent sections we sketch the proof of some of them
but for details we refer the reader to [LL].

We measure the energy in units of $mc^2$ where $m$
is the bare mass of the electron, the length in units of the Compton
wave length $ \ell_C = \hbar/mc$ of the bare electron.
We further choose $\ell_C^{-1} \sqrt{\hbar c}$ as the unit for the vector
potential $A$ and
$\ell_C^{-2} \sqrt{\hbar c}$ as the unit for the magnetic field $B$.
The argument is the dimensionless quantity $ \ell_C^{-1} x$. As usual,
$\alpha = e^2 / \hbar c \approx 1/137.04$ is the fine structure constant.

In the expression below, $A(x)$ denotes an ultraviolet cutoff radiation
field localised in a box $L\times L\times L$ with volume $V=L^3$,

\begin{equation}
A(x)=\frac {1 }{ \sqrt{2V}} \sum_{|k| < \Lambda} \sum_ {\lambda =1,2} \frac{1}{
\sqrt{|k|}}
\varepsilon_{\lambda}(k)\bigl[ a_{\lambda}(k) e^{ i x\cdot k} +
 a_{\lambda}^*(k) e^{- i x\cdot k} \bigr] \ .
\end{equation}
\label{beans}
The index $k=2 \pi n/L$ where $n \in \mathbb{Z}^3$, and the word cutoff refers
to the restriction to all values of $k$ with $|k| < \Lambda$.

The vectors $\varepsilon_{\lambda}(k)$ are the polarization vectors and are
normalized in such a way that
\begin{equation}
\varepsilon_{i}(k)\cdot \varepsilon_{j}(k) = \delta_{i,j}\ ,\
\varepsilon_{i}(k)\cdot k = 0 \ .
\end{equation}

The operators $a_{\lambda}(k)$ and $a_{\lambda}^{*}(k)$ satisfy the
commutation relation
\begin{equation}
\bigl[a_{\lambda}(k), a_{\lambda^{\prime}}^{*}(k^{\prime})\bigr]= 
\delta_{\lambda,\lambda^{\prime}} \delta(k,k^{\prime}) \ ,
\end{equation}
while all others commute with each other.

The energy of the radiation field can now be conveniently written as
\begin{equation}
H_f = \sum_ {|k| < \Lambda} \sum_{ \lambda =1,2}|k| a_{\lambda}^{*}(k)
a_{\lambda}(k) \ .
\end{equation}

These operators act on the Hilbert space generated by the polynomials
in $a_{\lambda}^{*}(k)$ acting on the vacuum $|0 \rangle$.

The self energy of (one or more) particles is the {\bf ground state
energy} of the Hamiltonian
\begin{equation}
H = {\rm kinetic \ energy} \ + \ H_f \ .
\end{equation}
where, as usual, the ground state energy of $H$ is defined to be 
\begin{equation}
E_0 = \inf_{\Psi} \frac{ \langle \Psi, H\ \Psi \rangle }{\langle \Psi, \Psi \rangle}\
.
\end{equation}

Typically, in the inquiry about the self--energy problem, i.e., 
the problem of computing
the self--energy for fixed, albeit small, $\alpha$  and for large $\Lambda$,
one proceeds via  perturbation theory.
First order perturbation theory will predict an energy
of the order of $\alpha \Lambda ^2$, and a higher order power counting
argument confirms the asymptotically large $\Lambda$ 
dependence of that calculation.
Our theorems below show that the predictions of perturbation theory
for the self--energy problem are wrong, if one is interested in the large  
$\Lambda$ asymptotics of the energy. If perturbation theory works
at all, then it works only for a range of $\alpha$ that vanishes as $\Lambda$
increases. 
In fact we deduce from the upper bound in Theorem \ref{theo:nonrel} 
that the size of this range shrinks at least as $\Lambda^{-2/5}$.

All the theorems below are asymptotic statements for large $\Lambda$ and 
for fixed $\alpha$. For actual bounds we refer the reader to [LL].
The first result concerns the self energy of a nonrelativistic electron
interacting with the radiation field. The Hamiltonian is given by

\begin{equation}
H= \frac{1}{2} (p+ \sqrt{\alpha} A(x))^2 +H_{f} \ ,
\label{bacon}
\end{equation}
where $p =-i\nabla$ and acts on $L^2(\mathbb{R}^3)\otimes {\mathcal F}$,
where ${\mathcal F}$ denotes the photon Fock space.

\begin{theorem}
The ground state energy, $E_0$, of the operator (\ref{bacon}) satisfies the 
bounds
\begin{equation}
C_1 \alpha^{1/2} \Lambda^{3/2} \ < \ E_0 \ <  \ C_2 
\alpha^{2/7} \Lambda^{12/7}
\label{eq:nonrel}
\end {equation}
\label{theo:nonrel}
\end{theorem}

We do not know how to get upper and lower bounds that are of the same
order in $\Lambda$, but we suspect that $\Lambda^{12/7}$ is the right
exponent.  This is supported by the following theorem in which the 
$p\cdot A$ term is omitted.

\begin{theorem} 
The ground state energy $E_0$ of the operator
\begin{equation}
\frac{1}{2}  \left[ p^2 + \alpha A(x)^2 \right] + H_f
\label{eq:2.28}
\end{equation}
satisfies the bounds
\begin{equation}
C_1  \alpha^{2/7} \Lambda ^{12/7}
\leq E_0 \leq C_2  \alpha^{2/7} \Lambda ^{12/7}
\label{eq:2.29}
\end{equation}
\label{theo:2.3}
\end{theorem}

While these results are not of direct physical relevance (since $E_0$ is
not observable), the many-body problem is of importance since it reveals a
dramtic difference between bosons and fermions.  

\begin{theorem}
The ground state
energy of $N$ bosons, 
$E^{boson}_0(N)$, with Hamiltonian
\begin{equation}
H(N)=\sum_{j=1}^N \frac{1}{ 2}(p_j +\sqrt \alpha A(x_j))^2  
+  H_f\
\label{manyham}
\end{equation}
satisfies the bounds 
\begin{equation}
 C_1 \sqrt{N}  
\sqrt{\alpha} \Lambda^{3/2} \leq
 E^{boson}_0(N) \leq
C_2 N^{5/7}\alpha ^{2/7} \Lambda^{12/7}
\label{eq:nonrelboson}
\end{equation}
\label{theo:nonrelboson}
\end{theorem}
Thus, the energy $E^{boson}_0(N)$ is not extensive, i.e., it costs a
huge energy to separate bosons. This has to be contrasted with the next
theorem about fermions.  The Hamiltonian is  the same as before but it
acts on  the Hilbert space \begin{equation} {\mathcal F}\otimes
\wedge_{j=1}^N L^2(\mathbb{R}^3; \mathbb{C}^2) ~ , \end{equation} where the
wedge product indicates the antisymmetric tensor product is taken.

\begin{theorem} 
The ground state energy, $E^{fermion}_0(N)$, of $N$ charged fermions
interacting with the radiation field satisfies
\begin{equation}
C_1 \alpha^{1/2} \Lambda^{3/2} N \ \leq \ E^{fermion}_0(N) \ \leq  \ C_2 
\alpha^{2/7} \Lambda^{12/7} N
\label{eq:nonrelfer}
\end {equation}
\label{theo:nonrelfer}
\end{theorem}

The {\bf ``relativistic'' kinetic energy} for an electron is
\begin{equation}
T^{\rm rel} = | p + \sqrt{\alpha} A (x) | = \sqrt{[p + \sqrt{\alpha} A(x)]^2}
\label{eq:4.1}
\end{equation}
with $p=-i\nabla$. (Really, we should take $\sqrt{[p+\sqrt{\alpha}A
(x)]^2 + 1}$, but since $x < \sqrt{x^2 + 1} < x + 1$, the difference
is bounded by $N$.)

\bigskip

Consider, first, the $N=1$ body problem with the Hamiltonian
\begin{equation}
H =T^{\rm rel} + H_f ~.
\label{eq:4.2}
\end{equation}

By simple length scaling (with a simultaneous scaling of the volume
$V$) we easily see that $E_0 = \inf {\rm{spec}}~ (H) = C \Lambda$. Our
goal here is 
to show that the constant, $C$, is strictly positive and to give an
effective 
lower bound for it. But we would like to do
more, namely investigate the dependence of this constant on
$\alpha$. We also want to show, later on, that for $N$ fermions the
energy is bounded below by a positive constant times $N \Lambda$. Our
proof will contain some novel --- even bizarre --- features. 

\begin{theorem}
For the Hamiltonian in \ref{eq:4.2} there are positive constants, $C, C',
C''$ such that
\begin{eqnarray*}
E_0 &\leq& C \sqrt{\alpha} \Lambda\\
E_0 &\geq& C' \sqrt{\alpha} \Lambda~{\rm{for~ small}}~ \alpha\\
E_0 &\geq& C'' \Lambda~{\rm{for ~ large}}~ \alpha ~ .
\end{eqnarray*}
\label{theo:4.1}
\end{theorem}

The generalization  of this to $N$ fermions is similar to the 
nonrelativistic generalization, except that the power of 
$\Lambda$ is the same on both sides of the inequalities.

\begin{theorem}
For $N$ fermions with Hamiltonian
$$H_N = \sum_{i=1}^N T^{\rm rel}(x_i) + H_f$$
there are positive constants $C, C', C'',$ independent of $\alpha$ and
$N$, such that
\begin{eqnarray}
E_0 &\leq& C N \sqrt{\alpha} \Lambda\nonumber   \\ % \notag \\
E_0 &\geq& C' N \sqrt{\alpha} \Lambda \quad {\rm{for~small}} ~ \alpha\nonumber 
 \\E_0 &\geq& C'' N \Lambda  \quad {\rm{for~large}} ~ \alpha
\label{eq:4.23}
\end{eqnarray}
\label{theo:4.3}
\end{theorem}

We close this introduction by mentioning one last result about the 
Pauli--operator.
The kinetic energy expression is given by
\begin{equation}
T^{\rm Pauli}=[\sigma \cdot (p+ \sqrt{\alpha}A(x))]^2 = (p+
\sqrt{\alpha}A(x))^2 + \sqrt{\alpha}\ \sigma \cdot B(x) \ .
\label{eq:4.4}
\end{equation}
where $\sigma$ denotes the vector consisting of the Pauli matrices. Observe that
this term automatically accounts for the spin--field interaction.
Our result for the self energy of a Pauli electron is the following.
\begin{theorem}
The ground state energy $E_0$ of the Hamiltonian with Pauli energy, 
\begin{equation}
\frac{1}{2}[\sigma \cdot (p+ \sqrt{\alpha}A(x))]^2 + H_f ,
\end{equation}
satisfies the bounds
\begin{eqnarray}
E_0 &\leq & C_3 \sqrt{\alpha}\Lambda^{3/2}  \\
E_0 &\geq & C_1 \alpha \Lambda  \quad
{\rm{for~small}} ~ \alpha\nonumber \\
E_0 &\geq & C_2 \alpha^{1/3} \Lambda  \quad
{\rm for~large} ~ \alpha \nonumber 
\end{eqnarray}
For $N$ fermions, the bounds above are multiplied by $N$ (and the
constants are changed).
\label{Pauli}
\end{theorem}
For the detail of the proof, we refer the reader to [LL]. We believe
that the upper bound is closer to the truth since the main
contributions to the self energy should come from the fluctuations of
the  $A^2$ term.

Theorem \ref{Pauli} has the following consequence for stability of
matter interacting with quantized fields. It was shown in \cite{LLS}
that a system of electrons and nuclei interacting with Coulomb forces,
with the Pauli kinetic energy for the electrons and with a {\it
classical} magnetic field energy is stable (i.e., the ground state
energy is bounded below by $N$) if and only if $\alpha$ is small
enough. In \cite{BFG},\cite{FFG} this result was extended to {\it
quantized}, ultraviolet cutoff magnetic fields (as here). Among other
things, it was shown in \cite{FFG} that the ground state energy, $E_0$,
of the electrons and nuclei problem is bounded below by $-\alpha^2
\Lambda N$ for small $\alpha$. Theorem \ref{Pauli} implies, as a
corollary, that for small $\alpha$ the {\it total energy} (including
Coulomb energies) is bounded below by $+\alpha \Lambda N$. In other
words, among the three components of energy (kinetic, field and
Coulomb), the first two overwhelm the third --- for small $\alpha$, at
least.

All of these statements are true without mass renormalization and the
situation could conceivably be more dramatic when the mass is
renormalized. In any case, the true physical questions concern energy
differences, and this question remains to be addressed.

%%%%%%%%%%%%%%%%%%%%%%%%%%%%%%%%%%%%%%%%%%%%%%%%%%%%%%%%%
\section{NON-RELATIVISTIC ENERGY BOUNDS}

{\it Theorem \ref{theo:nonrel}:} We sketch a proof of
Theorem \ref{theo:nonrel}.  It is clear by taking the state
$V^{-1/2}/ \otimes |0\rangle$ that the ground state energy is
bounded above by ${\rm{(const)}} \alpha \Lambda^2$, which is the same
result one gets from perturbation theory. Since the field energy in this
state vanishes, such a computation ignores the tradeoff between the
kinetic energy of the electron and the field energy. Thus, it is
important to quantify this tradeoff.  The main idea is to estimate the
field energy in terms of selected modes.

Consider the operators (field modes), parametrized by $y \in {\mathbb
R}^3$,  
\begin{equation}
L(y) = \frac{1}{\sqrt{2V}} \sum_{|k| < \Lambda, \lambda} \sqrt{|k|}
a_{\lambda}(k) \overline{v}_{\lambda} (k)e^{i k \cdot y} \ ,
\label{eq:2.4}
\end{equation}
with some arbitrary complex coefficients $v_{\lambda}(k)$.
The following lemma is elementary

\begin{lemma}
\begin{equation}
H_f \geq \int w(y) L^{*}(y) L(y) {\rm d} y\ 
\label{eq:2.5}
\end{equation}
provided that the functions $v_{\lambda}(k)$ and $w$ are chosen such that,
as matrices, 
\begin{equation}
|k|\delta_{\lambda^{\prime}, \lambda} \delta(k^{\prime}, k)
\geq  \frac{1}{2V} \overline{v}_{\lambda} (k) 
\widehat{w}(k-k^{\prime}) v_{\lambda^{'}} (k^{'}) \ ,
\label{eq:2.6}
\end{equation}
or equivalently, that
\begin{equation}
\sum_{|k| < \Lambda, \lambda} \frac{|f_{\lambda}(k)|^2}{|v_ {\lambda} (k)|^2}
 \geq \sum_{|k|,|k^{'}| < \Lambda, \lambda, \lambda ^{'}} \frac{1}{2V}
\overline{f_{\lambda}(k)}f_{\lambda^{\prime}} (k^{\prime})
\widehat{w}(k-k^{\prime})
\label{eq:2.7}
\end{equation}
for all $f_{\lambda}(k)$, where
$\widehat{w}(k) = \int e^{ik \cdot x} w(x) {\rm d} x$.
\label{lemma:2.2}
\end{lemma}  

For the proof, one simply notes that both sides of (\ref{eq:2.5}) are
quadratic 
forms in the creation and annihilation operators, and hence
(\ref{eq:2.6}) and (\ref{eq:2.7}) are 
necessary
and sufficient conditions for (\ref{eq:2.5}) to be true. \hfill{$\blacksquare$}

\bigskip

\begin{corollary}
\begin{equation}
H_f \geq -\frac{1}{2V} \sum_{|k| < \Lambda, \lambda} |k|
|v_{\lambda} (k)|^2 \ \int w(y) {\rm d} y + \frac{1}{4}
\begin{cases}
\int w(y) (L(y) + L^{*}(y))^2 {\rm d} y \\
 -\int w(y) (L(y) - \frac{1}{4} L^{*}(y))^2 {\rm d} y 
\end{cases}
\label{eq:2.8}
\end{equation}
\label{corollary:1}
\end{corollary}

To prove this, note that
\begin{equation}
L^{*} L = L L^{*}- \frac{1}{2V} \sum_{|k| < \Lambda , \lambda} 
|k| |v_{\lambda} (k)|^2 \ ,
\label{eq:2.9}
\end{equation}
and, quite generally for operators,
\begin{equation}
LL + L^{\ast} L^{\ast} \leq  L^{\ast} L + L L^{\ast} \ . ~~~~~~~~~~~~~~~\blacksquare
\label{eq:2.10}
\end{equation}
 
\bigskip

Returning to the proof of Theorem \ref{theo:nonrel} we start with the
lower bound. 
Denote by
\begin{equation}
\Pi(x) = \frac{-i}{\sqrt{2V}}\sum_{|k| < \Lambda , \lambda}
{\sqrt{|k|} \varepsilon_{\lambda}(k) } \left( a_{\lambda}(k) 
e^{ik \cdot x} - a^{*}_{\lambda}(k) e^{-i k \cdot x} \right) \ .
\label{eq:2.11}
\end{equation}
This operator is canonically conjugate to $A(x)$ in the sense that we
have the commutation relations
\begin{equation}
i[\Pi_i(x) , A_j(x)] = \delta_{i,j} \frac{1}{(3\pi)^2 } \Lambda ^3 \ .
\label{eq:2.12}
\end{equation}
For our calculation, it is important to note the
\begin{equation}
{\rm div} ~ \Pi(x) =0 \ .
\label{eq:2.13}
\end{equation}
Hence from (\ref{eq:2.12}) and (\ref{eq:2.13}) we get that (for all
$j$)
\begin{equation}
[p_j + \sqrt{\alpha} A_j(x) , \Pi_j(x)] = \sqrt{\alpha} \frac{i}{(3\pi)^2 } \Lambda ^3 \ . 
\label{eq:2.14}
\end{equation}
The inequality
\begin{equation}
\frac{1}{2 } (p+ \sqrt{\alpha}A(x))^2 + 2a^2 \Pi(x)^2 \geq
-ai \sum^{3}_{j=1}[p_j + \sqrt{\alpha} A_j(x) , \Pi_j(x)] \ ,
\label{eq:2.15}
\end{equation}
valid for all positive numbers $a$, yields
\begin{equation}
\frac{1}{2} (p+ \sqrt{\alpha} A(x))^2 + H_f 
\geq a \sqrt{\alpha} \frac{1}{(3 \pi)^2} \Lambda ^3 + H_f - 2a^2 \Pi(x)^2 \ .
\label{eq:2.16}
\end{equation}
Now, with
\begin{equation}
v_ {\lambda} (k) = (3 \pi)\frac{\varepsilon_{\lambda}(k)}{\Lambda ^{3/2}}
\label{eq:2.17}
\end{equation}
and
\begin{equation}
w(y) = \delta (x-y) \ ,
\label{eq:2.18}
\end{equation}
it is elementary to see that (\ref{eq:2.7}) is satisfied.
Hence Corollary \ref{corollary:1} yields
\begin{equation}
H_f \geq \frac{9 \pi^2}{4\Lambda ^3} \Pi(x)^2 - \frac{9}{8} \Lambda
\label{eq:2.19}
\end{equation}
Choosing $a= (3 \pi)/(\sqrt{2} \Lambda^{3/2})$ 
yields the {\it lower bound}
\begin{equation}
H \geq \frac{1}{3 \pi}\sqrt{\frac{\alpha}{2}} \Lambda ^{3/2} - 
\frac{9}{8}\Lambda
\label{eq:2.20}
\end{equation}

\bigskip
The idea of using a commutator, as in (\ref{eq:2.15}), (\ref{eq:2.16})
to estimate the ground state energy, goes back to the study of the
polaron \cite{LY}.

For the upper bound we take a simple trial function of the form
\begin{equation}
\phi(x) \otimes \Psi
\label{eq:2.21}
\end{equation}
where $\Psi \in {\mathcal F}$  is normalized and $\phi(x)$ is a {\it real} function
normalized in
$L^2({\mathbb R}^3)$. An upper bound to the energy is thus given by
\begin{equation}
\frac{1}{2} \int|\nabla \phi(x)|^2 {\rm d} x +\frac{ \alpha}{2}
\int \phi(x)^2 \left(\Psi, A(x)^2 \Psi\right) {\rm d} x + (\Psi,H_f
\Psi) ~ .
\label{eq:2.22}
\end{equation}
It is not very difficult to see that the last two terms can be concatenated
into the following expression.
\begin{equation}
\frac{1}{2}\int \left(\Psi, \left[ \Pi(x)^2 +\alpha  A(x) (-\Delta
+\phi(x)^2) A(x)\right] \Psi\right) {\rm d}x ~ -\frac{1}{2} {\rm{Tr}}
\sqrt{P-\Delta P} \ .
\label{eq:2.23}
\end{equation}
Here, $P$ is the projection onto the divergence free vector fields with
ultraviolet cutoff $\Lambda$. This can be deduced by writing the field
energy in terms of $\Pi(x)$ and $A(x)$. The first term in (\ref{eq:2.23}) is
a sum of harmonic oscillators whose zero point energy is given by
\begin{equation}
\frac{1}{2} {\rm{Tr}} \sqrt{ P\left(-\Delta + \alpha 
\phi(x)^2 \right)P} 
\label{eq:2.24}
\end{equation}
and hence 
\begin{equation}
\frac{1}{2} {\rm{Tr}} \sqrt{ P\left(-\Delta + \alpha 
\phi(x)^2 \right)P} -\frac{1}{2} {\rm{Tr}} \sqrt{P (-\Delta) P}\ ,
\label{eq:2.25}
\end{equation}
is an exact expression for the ground state energy.
Using the operator monotonicity of the square root we get as an upper bound on
(\ref{eq:2.23})
\begin{equation}
\frac{1}{2} \int|\nabla \phi(x)|^2 {\rm d}x + \frac{1}{2} \sqrt{\alpha} 
\ {\rm{Tr}}\sqrt{ P \phi(x)^2 P} \ .
\label{eq:2.26}
\end{equation}
As a trial function we use the positive function
\begin{equation}
\phi(x) = {\rm{const.}} K^{-3/2} \int \left(1 - \frac{|k|}{K} \right)^3_+ e^{ik \cdot
x} {\rm d} x \ .
\label{eq:2.27}
\end{equation}
Optimizing the resulting expression over $K$ yields the stated result.
For details we refer the reader to \cite{LL}.  \hfill{$\blacksquare$}

\bigskip

It is natural to ask, how good this upper bound is. If we neglect the
cross terms in $(p+A)^2$, i.e., we  replace the kinetic energy by $p^2 +
\alpha A(x)^2$, then we have Theorem \ref{theo:2.3}, which we prove next.

{\it Theorem \ref{theo:2.3}:} The upper bound was already given in
Theorem \ref{theo:nonrel} because $< p\cdot A > = 0$ in the state
(\ref{eq:2.21}).  Loosely speaking equation (\ref{eq:2.12}) expresses the
Heisenberg uncertainty principle for the field operators. An uncertainty
principle that is quite a bit more useful is the following.

\begin{lemma}
 The following inequality holds in the sense of quadratic forms
\begin{equation}
\Pi(x)^2 \geq \frac{1}{4}\frac{1}{(3\pi)^4}\Lambda^6 \frac{1}{A(x)^2}\ .
\label{eq:2.30}
\end{equation}
\label{lemma:2.4}
\end{lemma}

For the proof note that $[A_j(x), A_k(y)] = 0$ and compute
\begin{equation}
i[\Pi(x)_j, \frac{A_j(x)}{A(x)^2}]
=\frac{1}{(3\pi)^2}\Lambda^3 \left[\frac{1}{A(x)^2} -
2 \left( \frac{A_j(x)}{A(x)^2}\right)^2 \right] \ ,
\label{eq:2.31}
\end{equation}
and summing over $j$ we obtain that
\begin{equation}
i \sum_{j=1}^3[\Pi(x)_j, \frac{A_j(x)}{A(x)^2}]
=\frac{1 }{(3\pi)^2}\Lambda^3 \frac{1}{A(x)^2}
\label{eq:2.32}
\end{equation}

Our statement follows from the Schwarz inequality. \hfill{$\blacksquare$}

\bigskip
To prove Theorem \ref{theo:2.3} we return to Lemma \ref{lemma:2.2} and
choose $v_{\lambda}(k)= 
\varepsilon_{\lambda}(k)$ and $w(x)$ any function $\leq 1$. 
Corollary \ref{corollary:1} applied to each of the 3 components of $\Pi(x)$ then yields 
\begin{equation}
H_f \geq \frac{ 1}{4} \int w(x-y) \Pi (y)^2 {\rm d} y - \Lambda^4 
\frac{3}{8 \pi^2} \int w(y) {\rm d} y \ ,
\label{eq:2.33}
\end{equation}
for every $x \in {\mathbb R}^3$.
By Lemma \ref{lemma:2.2} the right side is bounded below by
\begin{equation}
\Lambda^6 \int w(x-y) \frac{1}{A(y)^2} {\rm d} y - 
\Lambda^4\int w(y) {\rm d} y \ ,
\label{eq:2.34}
\end{equation}
and hence
\begin{eqnarray}
\langle \Psi, H \Psi \rangle  &\geq&  \frac{1}{2} \int \langle \nabla
\Psi(x), \nabla \Psi(x) 
\rangle {\rm d} x + \frac{\alpha}{2} \int \langle \Psi(x), A(x)^2
\Psi(x) \rangle {\rm d} x     \nonumber  \\
&+&\Lambda^6 \int w(x-y) \langle \Psi(y),
\frac{1}{A(x)^2}\Psi(y) \rangle {\rm d}
y {\rm d} x   \nonumber  \\
&-& \Lambda^4\int w(y) {\rm d} y \int \langle \Psi(x), \Psi(x) 
\rangle {\rm d} x \ .
\label{eq:2.35}
\end{eqnarray}
By Schwarz's inequality the second and third term together are bounded
below by
\begin{equation}
\sqrt{\frac{\alpha}{2}} \Lambda^3\int \langle \Psi(x), \Psi(y) 
\rangle \frac{w(x-y)}{\sqrt{\int w(z)
{\rm d}z}} {\rm d} x {\rm d} y \ .
\label{eq:2.36}
\end{equation} 
If we restate our bound in terms of Fourier space variables we get
\begin{equation}
\int \left[\frac{|p|}{2}^2 + \sqrt{\frac{\alpha}{2}} ~ \Lambda^3
\frac{\widehat{w}(p)}{\sqrt{ \widehat{w}(0)}} 
\right] \langle \widehat{\Psi}(p), \widehat{\Psi}(p)\rangle {\rm d} p
 -\Lambda^4 \widehat{w}(0) \int \langle \widehat{\Psi}(p),
\widehat{\Psi}(p) \rangle {\rm d}p \ .
\label{eq:2.37}
\end{equation}
Choosing the function $\widehat{w}(p)$ to be $(2 \pi)^3 
\Lambda^{-18/7}$ times the 
characteristic function of the ball of radius $\Lambda^{6/7}$, we have that
$w(x)
\leq 1$ and it remains to optimize (\ref{eq:2.37}) over all normalized states
$\widehat{\Psi}(p)$. This is easily achieved by noting that the function 
\begin{equation}
\frac{1}{2} |p|^2 + \sqrt{\frac{\alpha}{ 2}} \Lambda^3 \frac{\widehat{w}(p)}{\sqrt{ \widehat{w}(0)}} 
\label{eq:2.38}
\end{equation}
is everywhere larger than $\Lambda ^{12/7}$. 
Strictly speaking, the function $w(x)$ should be positive in order for
the argument that led to (\ref{eq:2.36}) to be valid. This can be
achieved with a 
different choice of $w(x)$ that is more complicated but does not
change the argument in an essential way.

\section{NON-RELATIVISTIC MANY-BODY ENERGIES}
\label{non-relativistic}

A problem that has to be addressed is the energy of $N$ particles
(bosons or fermions) interacting with the radiation field. If $E_0 =
E_0(1)$ is the energy of one particle (which we estimated in the
preceding section) then, ideally
the energy, $E_0(N)$, of $N$ particles (which trivially satisfies $E_N
\leq NE$, since the $N$ particles can be placed infinitely far apart)
ought to be, {\it exactly},
\begin{equation}
E_0(N) = NE_0 
\end{equation}
in a correct QED.
In other words, in the absence of nuclei and Coulomb potentials, there
should be {\it no binding} caused by the field energy $H_f$. This is
what we seem to observe experimentally, but this important topic does
not seem to have been discussed in the QED literature.

Normally, one should expect binding, for the following mathematical
reason: The first particle generates a field, $A(x)$, and energy
$E_0$. The second particle can either try to generate a field
$A(x+y)$, located very far away at $y$ or the second particle can try
to take advantage of the field $A(x)$, already generated by the first
particle, and achieve an insertion energy lower than $E_0$.

Indeed, this second phenomenon happens for bosons --- as expected. For
fermions, however, the Paul principle plays a crucial role (even in in
the absence of Coulomb attractions). We show that $E_0(N) \geq C N
E_0$ for fermions, but we are unable to show that the universal
constant $C = 1$. Even if $C < 1$, the situation could still be saved
by mass renormalization, which drives the bare mass to zero as
$\Lambda$ increases, thereby pushing particles apart.

\subsection{BOSONS}
\label{bosons}

{\it Theorem \ref{theo:nonrelboson};} This theorem concerns the ground
state energy of $N$ charged bosons. the Hamiltonian is given by
\ref{manyham} acting on the Hilbert space of symmetric functions tensored
with the photon Fock space $\mathcal F$. It states, basically, that
$C_1 \sqrt{N} \sqrt{\alpha} \Lambda^{3/2} \leq 
E_0^{boson}(N) \leq  C_2 N^{5/7}  \alpha^{2/7} \Lambda^{12/7}$.

The proof follows essentially that of the one particle case. The
interesting fact is that it implies {\it binding} of {\it charged
bosons} (in the absence of the Coulomb repulsion). The binding energy
is defined by 
$$
\Delta E (N) = E_0 (N) - NE_0(1)
$$
and satisfies the bounds
\begin{eqnarray}
\Delta E(N) &\geq& C_1 \sqrt{N} \sqrt{\alpha}~ \Lambda^{3/2} -
C_2 N  \alpha^{2/7} \Lambda^{12/7}  \nonumber \\
\Delta E(N) &\leq& C_2 N^{2/7} 
\alpha^{2/7} \Lambda^{12/7} -  C_1 N \sqrt{\alpha}~ \Lambda^{3/2}
\label{eq:3.6}
\end{eqnarray}
which can be made negative for appropriately chosen $N$ and
$\Lambda$. 
There will be binding for all large
enough $N$, irrespective of the cutoff $\Lambda$. It also has to be
remarked that the Coulomb repulsion will, in all likelihood, not alter
this result since it has an effect on energy scales of the order of
$\Lambda$ and not $\Lambda^{12/7}$ or $\Lambda^{3/2}$.

\subsection{FERMIONS}
\label{fermions}

The real issue for physics is what happens with fermions. We cannot show
that there is no binding but we can show that the energy is extensive
as in Theorem \ref{theo:nonrelfer}.
The Hamiltonian is the same as (\ref{manyham}) but it acts on
antisymmetric functions tensored with ${\mathcal F}$. (Spin can be
ignored for present purposes.)

\noindent {\it Rough sketch of the proof of Theorem \ref{manyham}}.

The difficulty in proving this theorem stems from the fact that the
field energy is not extensive in any obvious way.

Define $\underbar{X} = (x_1, \cdots , x_N)$ and define the function
$$
n_j (\underbar{X}, R) = \# \{ x_i \neq x_j ~:~ |x_i - x_j| < R \} ~ .
$$
This function counts the number of electrons that are within a
distance $R$ of the $j^{\rm{th}}$ electron.
Note that this function is not smooth, so that all the following
computations have to be modified. (See \cite{LL}.) 
We save half of the kinetic energy and write
$$  %\begin{equation*}
H = \frac{1}{4} \sum^{N}_{j=1} (p_j + \sqrt{\alpha} A (x_j))^2 +
H' ~ .
$$   %\end{equation*}
We apply the commutator estimate (\ref{eq:2.14}) to the pair
$$
i [p_j + \sqrt{\alpha} A (x_j), \frac{1}{\sqrt{N_j (\underbar{X}, R) +
1}} \Pi (x_j)]
$$
and obtain the bound (with the caveat mentioned above), for all
$\alpha > 0$,
\begin{equation}
H' \geq a \sqrt{\alpha} \Lambda^3 \sum^N_{j=1}
\frac{1}{\sqrt{N_j(\underbar{X}, R) + 1}} - a^2 \sum^N_{j=1} \frac{1}{N_j
(\underbar{X}, R) + 1} {\rm{F}} (x_j)^2 + H_f
\label{eq:3.8}  %losseq
\end{equation}
The next two steps are somewhat nontrivial and we refer the reader to
\cite{LL}. First one notes that the modes $F(x_i)$ and $F(x_j)$ are essentially
orthogonal (i.e., they commute) if $|x_i- x_j| > \Lambda^{-1}~ .$
Ignoring the technical details of how this is implemented, the key 
observation is that the last two terms in (\ref{eq:3.8}) can be
estimated from 
below by $ - N\Lambda$ provided $a = \Lambda^{-3/2}$.

The next ingredient is a new Lieb-Thirring type estimate involving the
function $N_j(\underbar{X},R)$. It is here and only here that the
Pauli exclusion 
principle is invoked.

\begin{theorem}
On the space $\wedge^{N}_{j=1} L^2 {\mathbb R}^3; {\mathbb C}^q)$ of
antisymmetric functions
\begin{equation}
\sum_{j=1}^{N} (p_j + \sqrt{\alpha} A (x_j))^2 \geq \frac{C}{q^{2/3}}
\frac{1}{R^2} \sum^N_{j=1} N_j (\underbar{X}, R)^{2/3}
\end{equation}
with $C \geq 0.00127$. An analogous inequality holds for the
relativistic case as well:
\begin{equation}
\sum_{j=1}^{N} |p_j + \sqrt{\alpha} A (x_j)| \geq \frac{C}{q^{1/3}}
\frac{1}{R} \sum^N_{j=1} N_j (\underbar{X}, R)^{1/3}
\end{equation}
\label{theo:3.3}
\end{theorem}

By using the kinetic energy previously saved together with
(\ref{eq:3.8}) and the previous discussion, we get
$$    %\begin{equation*}
H \geq \sum^N_{j=1} \left\{  N_j (\underbar{X},R)^{2/3}
+  \sqrt{\alpha} \Lambda^{3/2} \frac{1}{\sqrt{N_j
(\underbar{X},R)+1}} \right\} - N \Lambda ~ .
$$       %\end{equation*}

By minimizing over $N_j$ the desired estimate is obtained. The upper
bound is fairly elementary and is omitted. \hfill $\blacksquare$
%%%%%%%%%%%%%%%%%%%%%%%%%%%%%%%%%%%

\section{RELATIVISTIC ENERGY BOUNDS}
\label{relativistic}

{\it Theorem \ref{theo:4.1}:} Sketch of Proof.

An upper bound for $E_0$ is easy to obtain, but it is indirect. Note
that
\begin{equation}
|p + \sqrt{\alpha} A(x) | \leq \varepsilon [ p + \sqrt{\alpha} A (x)
 ]^2 + (4 \varepsilon)^{-1}
\label{eq:4.3}
\end{equation}
for any $\varepsilon > 0$. Take $\Psi = f (x) \otimes | 0\rangle$ with
$|0\rangle$ being the Fock space vacuum. Using (\ref{eq:4.3}) 

\begin{eqnarray}
(\Psi, H \Psi) &\leq& \varepsilon ~\int_{\mathbb{R}^{\rm{3}}} \{ \alpha
\langle 0 | A 
(x)^2 | 0 \rangle | f (x)|^2 + | \nabla f (x)|^2 \} dx +
\varepsilon^{-1} \nonumber \\
&=& \frac{\varepsilon \alpha \Lambda^2}{4 \pi} ~ + \int |\nabla f|^2 +
\frac{1}{4 \varepsilon}~,
\label{eq:4.5a}
\end{eqnarray}
since $\langle 0 | A (x)^2 | 0 \rangle = (2 \pi)^{-3} \int_{|k|<
\Lambda} |k|^{-1} 
dk = \Lambda^2/4\pi^2$. We can now let $f(x) \rightarrow
V^{-\frac{1}{2}}$ and take $\varepsilon = (\pi/ \alpha)^{1/2}
\Lambda^{-1}$, whence
\begin{equation}
E_0 \leq (\alpha/4 \pi)^{1/2} \Lambda ~.
\label{eq:4.5b}
\end{equation}

Now we turn to the lower bound for $H$.

{\bf Step 1}: Since $x \rightarrow \sqrt{x}$ is operator monotone,
\begin{equation}
T > T_1 = | p_1 + \sqrt{\alpha} A_1 (x) | ~,
\label{eq:4.6}
\end{equation}
where the subscript 1 denotes the 1 component (i.e., the
$x$-component) of a vector. By replacing $T$ by $T_1$, we are now in a
position to remove $A_1$ by a gauge transformation - but it has to be
an {\it operator-valued gauge transformation}. The use of such a gauge
transformation is is a novelty,
as far as we are aware, in QED.

To effect the gauge transformation, set
\begin{equation}
\varphi (x) = \frac{1}{\sqrt{2V}}~ \sum_{h,\lambda} ~
\frac{\varepsilon^{\lambda}(k)_1}{\sqrt{(k)}} ~ \left[ a_{\lambda} (k)
+ a_{\lambda}^{\ast} (-k) \right] \frac{e^{ik_1 x_1} -1}{ik_1} ~
e^{ik_{\perp}x_{\perp}}
\label{eq:4.7}
\end{equation}
with $x_{\perp} = (x_2, x_3$). Then $[A_j (x), \varphi (x) ] = 0, ~~j= 1,2,3$
and $p_1 \exp~ [i \varphi (x)] = -A_1(x)$. The unitary $U = \exp~ [i
\varphi (x)]$ is a gauge transformation, but it is {\it
operator-valued}. We 
have
\begin{eqnarray}
U^{-1} | p_1 + A_1 (x) | U &=& |p_1| \nonumber \\ 
U^{-1} a_{\lambda} (k) U &=& a_{\lambda}(k) + f_{\lambda} (k, x)
\nonumber \\
U^{-1} a_{\lambda}^{\ast} (k) U &=& a_{\lambda}^{\ast}(k) +
\bar{f}_{\lambda} (k, x) \nonumber \\ 
\stackrel{\sim}H_f ~=~ U^{-1} H_f U &=& \sum_{k,\lambda} |k| 
[ a_{\lambda}^{\ast} (k) + \bar{f}_{\lambda} (k,x) ] 
[ a_{\lambda} (k) + f_{\lambda} (k,x) ]
\label{eq:4.8}
\end{eqnarray}
with
\begin{equation}
f_{\lambda} (k,x) ~=~ \sqrt{\frac{\alpha}{2V}} ~ \sum_{k,\lambda} ~
\frac{\varepsilon_{\lambda}(k)_1}{|k|} ~\frac{e^{-i k_1 x_1}-1}{k_1} ~
e^{-ik_{\perp}x_{\perp}} ~ .
\label{eq:4.9}
\end{equation}

Since $p_{\perp}$ does not appear in our new Hamiltonian,
\begin{equation}
\stackrel{\sim}H ~=~ U^{-1} HU ~=~ |p_1|+~\stackrel{\sim}H_f ~,
\label{eq:4.10}
\end{equation}
the variable $x_{\perp}$ appears only as a parameter, and thus we can set
$x_{\perp}$ = constant = (0,0), by translation invariance, and replace
$\mathbb{R}^3$ by $\mathbb{R}^1 = \mathbb{R}$.

{}From now on $x_1 = x$ and, $p_1 = p = -i ~ d/dx$.

{\bf Step 2}: The dependence on $x$ now appears in
$\stackrel{\sim}H_f$ instead of in 
the kinetic energy, $|p|$. For each $x$ we can try to put
$\stackrel{\sim}H_f$ into 
its ground state, which is that of a displaced harmonic
oscillator. But, since this state depends on $x$, to do so will
require a great deal of kinetic energy, $|p_1|$.

Let $\Psi$ be a normalized wave-function, i.e., a function on
$L^2(\mathbb R) \otimes {\mathcal{F}}$. We write it as $\psi_x$ where
$\psi_x \in {\mathcal{F}}$. Thus, with $\langle \cdot~ ,~ \cdot \rangle$
denoting the 
inner product on ${\mathcal{F}}, \int_{\mathbb R} \langle \psi_x, \psi_x
\rangle dx = 1$.

Decompose $\mathbb R$ as the disjoint union of intervals of length
$\ell/\Lambda$, where $\ell$ is a parameter to be determined
later. Denote these intervals by $I_j,~ j = 1, 2, \ldots~.$ A simple
Poincar\'e type inequality gives, for $g ~:~ L^2 (\mathbb R)
\rightarrow 
{\mathbb{C}}$, 
$$(g,|p| g) \geq C_1 \frac{\Lambda}{\ell} \sum_j 
\int_{I_j} \{ | g(x) |^2 - |\bar{g}_j |^2 \} dx ~ ,
$$
where $\bar{g}_j = \frac{\Lambda}{\ell} \int_{I_j} g (x) dx$ is the average
of $g$ in $I_j$. Then
\begin{equation}
(\Psi, |p| \Psi) ~ \geq C_1 \frac{\Lambda}{\ell} ~\sum_j ~
\int_{I_j} ~ \{ \langle \psi_x, \psi_x \rangle ~-~ \langle
\bar{\psi}_j, \bar{\psi}_j \rangle~\} dx ~ .
\label{eq:4.11}
\end{equation} 

{\bf Step 3}: Next, we analyze $\stackrel{\sim}H_f$. We think of this
as an operator 
on ${\mathcal{F}}$, parameterized by $x \in \mathbb R$.
We would like $\stackrel{\sim}H_f$ to have a gap so we 
define
\begin{equation}
H_x ~=~ \frac{\Lambda}{2} \sum_{\varepsilon \Lambda
\leq |k| \leq \Lambda} ~ \sum_{\lambda} ~ [a^+_{\lambda} (k) +
\bar{f}_{\lambda} (k,x)]~\cdot~ [~h.c.~]
\label{eq:4.12}
\end{equation}

Clearly, $\stackrel{\sim}H_f \geq H_x$ and
\begin{equation}
(\Psi, \stackrel{\sim}H \Psi) \geq \frac{\Lambda}{\ell} \sum_{j}
\int_{I_j} \langle \psi_x, \psi_x \rangle - \langle \bar{\psi}_j,
\bar{\psi}_j \rangle + \langle \psi_x, H_x \psi_x
\rangle dx ~ .
\label{eq:4.13}
\end{equation}
{}For each interval $I_j$ we can minimize (\ref{eq:4.13}) subject to
$\int_{I_j}\langle \psi_x, \psi_x \rangle dx$ fixed. This leads to
\begin{equation}
(h_j ~ \psi)_x = e ~ \psi_x
\label{eigenvalue1}
\end{equation}
with
\begin{equation}
(h_j ~ \psi)_x = \frac{\Lambda}{\ell} ~ \psi_x -
\frac{\Lambda}{\ell} ~ \bar{\psi}_j + H_x ~ \psi_x
\label{eigenvalue2}
\end{equation}
Obviously, this eigenvalue problem (\ref{eigenvalue1}, \ref{eigenvalue2}) is the same for all
intervals $I_j$, so we shall drop the subscript $j$ and try to find
the minimum $e$.

A lower bound to $h_j$ (and hence to $e$) can be found
by replacing $H_x$ by 
$$\widehat{H}_x = \frac{\Lambda}{2} ( 1 - \Pi_x) ~,
$$
where $\Pi_x = |g_x\rangle \langle g_x|$ is the projector onto the
ground state, $|g_x\rangle,~ {\rm{of}}~ H_x$.

If we substitute $\widehat{H}_x$ into (\ref{eigenvalue2}) the corresponding eigenvalue
equation (\ref{eigenvalue1}) becomes soluble. Multiply (\ref{eigenvalue1}) on the left by
$\langle g_x|$, whence

\begin{equation}
\left( \frac{\Lambda}{\ell} - e \right) \langle g_x, \psi_x \rangle =
\frac{\Lambda}{\ell} \langle g_x, \bar{\psi} \rangle
\label{eq:4.16}
\end{equation}
Then, substitute (\ref{eq:4.16}) into (\ref{eigenvalue2}) and integrate
$\int_I dx$ to find 

\begin{equation}
\frac{1}{2} \Lambda^3 \ell^{-2} \left( \int \Pi_x dx \right)
\bar{\psi} = \left ( \frac{\Lambda}{\ell} - e \right)
\left(\frac{\Lambda}{2} - e \right) \bar{\psi} ~ . 
\label{eq:4.17}
\end{equation}

We know that $e < \Lambda/2$ because we could take $\psi_x$ =
constant as a trial function, and then use $\stackrel{\sim}H_x \leq
\Lambda/2$. Also, $e < \Lambda/\ell$, because we could take 
$\Psi = \delta_{x_{0}} | g_{x_{0}} \rangle ~ .$

{\bf Step 4}: Eq. (\ref{eq:4.17}) will give us a lower bound to $e$ if
we can 
find an upper bound to $Y = (\Lambda/\ell) \int_I \Pi_x dx ~.$ To do
this note that
\begin{eqnarray}
Y^2 &\leq& {\rm{Trace}} ~ Y^2 = \left( \frac{\Lambda}{\ell} \right)^2
\int_I \int_I |\langle g_x, g_y \rangle |^2 dxdy\nonumber \\ 
&=& \left( \frac{\Lambda}{\ell} \right)^2 \int_I\int_I \exp \{ -
\frac{\alpha}{2V} \sum_{\varepsilon \Lambda \leq |k| \leq \Lambda}
\sum_{\lambda}|f_{\lambda} (k,x) - f_{\lambda} (k,y)|^2 dxdy \}
\label{eq:4.18}
\end{eqnarray}
Noting that $\sum_{\lambda = 1}^{2} e_{\lambda}(k)_1^2 = 1 -
k_1^2/k^2$, the quantity \{ \quad \} in (\ref{eq:4.18}) becomes (as $V
\rightarrow \infty$)

\begin{equation}
\{ \quad \} = -\frac{2 \alpha}{(2 \pi)^3} \int_{\Lambda/2 < |k|
< \Lambda} \frac{1}{|k|^3 k_1^2} (k_{\perp}^2) [ \sin
\frac{k_1}{2} (x-y) ]
\label{eq:4.19}
\end{equation}

After some algebra we find that

\begin{equation}
\left( \frac{1}{\ell} - \frac{e}{\Lambda} \right) \left( \frac{1}{2} -
\frac{e}{\Lambda} \right) \leq \frac{1}{2 \ell} ({\rm{Trace}}~
Y^2)^{1/2} \leq \frac{1}{2 \ell} \sqrt{K_{\ell} (\alpha)}
\label{eq:4.20}
\end{equation}
where
\begin{eqnarray}
K_{\ell} (\alpha) &=& \int_0^1 \int_0^1 \exp \left[ - \alpha
\frac{\ell}{\pi^2} |x-y| \int_0^{|x-y|\ell/4} \left( \frac{\sin t}{t}
\right)^2 ~ dt \right] ~ dxdy ~ .\nonumber   \\
&\leq& \int_{-1/2}^{1/2} \exp [ - \alpha x^2 \ell^2 / 8 \pi ] ~ dx ~. 
\label{eq:4.21}
\end{eqnarray}
We find that
\begin{eqnarray}
K_{\ell} (\alpha) &\sim& 1 - \alpha \ell^2 / 96 \pi, \quad \ell^2 \alpha
~ {\rm{small}}\\
&\sim& \sqrt{2} \pi ( \alpha \ell^2)^{1/2}, \quad \ell^2 \alpha ~
{\rm{large}} \nonumber
\label{eq:4.22}
\end{eqnarray}
If $\alpha$ is small we take $\ell \sim \alpha^{-1/2}$.
If $\alpha$ is large we
take $\ell = 2$. This establishes our theorem for $N=1$. \hfill{$\blacksquare$}

\smallskip

{\it Theorem \ref{theo:4.3}:} Sketch of Proof.

For $N > 1$ we can decompose $\mathbb{R}^3$ into cubic boxes $B_j,
j=1,2,3, \ldots$ of size $\ell \Lambda$ and ``borrow'' $\frac{1}{2}
|p+A(x)|^2$ kinetic energy from each particle. That is, $H_N =
H_N^{1/2} + \frac{1}{2} T_N$ with $T_N = \sum_{i=1}^{N} T(x_i)$. The
Pauli principle will then yield an energy for $\frac{1}{2}T_N$ that is
bounded below by (const.) $(n_{j-1})^{4/3}$, where $n_j$
is the particle number in box $B_j$.

\bibliographystyle{amsalpha}

\end{document}